\documentclass[11pt,a4paper]{article}
\usepackage{amsmath,amssymb,amsfonts}
\usepackage{graphicx}
\usepackage{xcolor}
\usepackage[lmargin=25mm,rmargin=25mm,bmargin=25mm,tmargin=25mm]{geometry}
\usepackage[showseconds=false]{datetime2}
\usepackage[bibencoding=utf8,sorting=nyt,hyperref=true,backend=biber]{biblatex}

\usepackage[colorlinks,verbose=false,linkcolor=blue,anchorcolor=blue,citecolor=blue,urlcolor=blue,pdfauthor={Keith Briggs and Ibrahim Nur},pdfmenubar=false,pdffitwindow=true,pdfwindowui=false,pdftitle={CRRM: A 5G system-level simulator},pdfkeywords={},pdfsubject={},plainpages=false]{hyperref} 

\def\BibTeX{{\rm B\kern-.05em{\sc i\kern-.025em b}\kern-.08em
    T\kern-.1667em\lower.7ex\hbox{E}\kern-.125emX}}

\def\cpp{{C\hspace{-.05em}\raisebox{.4ex}{\tiny\bf ++}}}

\graphicspath{{./}{../img/}} 
\bibliography{%
references,%
}

\begin{document}
\title{\vspace{-30mm}CRRM: A 5G system-level simulator}
\author{Keith Briggs \&\ Ibrahim Nur}
\date{\the\year-\ifnum\month<10\relax0\fi\the\month-\ifnum\day<10\relax0\fi\the\day\;\;\DTMcurrenttime}

\maketitle

\begin{abstract}
\noindent{}System-level simulation is indispensable for developing and testing
novel algorithms for 5G and future wireless networks, yet a gap persists
between the needs of the machine learning research community and the available
tooling. To address this, we introduce the Cellular Radio Reference Model
(CRRM), an open-source, pure Python simulator we designed specifically for
speed, usability, and direct integration with modern AI frameworks. The core
scientific contribution of CRRM lies in its architecture, which departs from
traditional discrete-event simulation. We model the system as a set of
inter-dependent computational blocks which form nodes in a directed graph. This
enables a compute-on-demand mechanism we term \textit{smart update}. 


\end{abstract}

\section{Introduction}

The optimization of the performance of modern cellular networks increasingly
depends on machine-learning algorithms, and these require simulators which
integrate directly with AI frameworks based on Python, such as PyTorch and
TensorFlow. A significant usability gap exists, as established high-fidelity
simulators like \texttt{ns-3} are built on \cpp\ and demand substantial,
specialised expertise, creating a barrier to entry for researchers focused on
AI applications. This project was initiated to address this specific gap by
developing the Cellular Radio Reference Model (CRRM), a pure Python,
open-source 5G system-level simulator. Its performance is founded on a
compute-on-demand architecture that avoids redundant calculations; timing tests
demonstrate this \textit{smart update} mechanism delivers a speed-up factor of
close to 2, over a full system recalculation in scenarios with a 10\% UE
mobility factor.


The simulator's utility is not confined to its architecture. Its capabilities
are grounded in a library of validated 3GPP TR 38.901 propagation models,
including Rural Macrocell (RMa), Urban Macrocell (UMa), Urban Microcell (UMi),
and Indoor Hotspot (InH), which allows for the analysis of a diverse range of
scenarios. The physical layer modelling extends to advanced features, such as
3-sector antenna patterns that are based on 3GPP specifications, which creates
a clear angular dependency on UE throughput. Furthermore, CRRM is a functional
tool for investigating resource management strategies. It implements
subband-based interference coordination, where a UE's
Signal-to-Interference-plus-Noise Ratio (SINR) can be improved from 0dB to
20dB through cell power coordination, and a tunable resource allocation
fairness parameter ($p$), to study throughput distribution among users.

This paper documents the design and implementation of the CRRM simulator. It
begins by detailing the compute-on-demand architecture that underpins its
performance. It then presents the suite of implemented physical layer models
and network features. Finally, it provides evidence of the simulator's
scientific validity by comparing its output against known analytical theory for
a Poisson Point Process network deployment (CRRM example 12), and quantifies the
performance benefits of its core design.

\section{Core architecture}

The fundamental design philosophy of CRRM is a dependency chain of
computational blocks, an architectural choice which represents a specific
hypothesis about the nature of system-level cellular simulation. This paradigm
deviates from the discrete-event schedulers that form the core of traditional
network simulators like \texttt{ns-3}. The implicit argument is that for this
problem domain, where the system state is largely static except for discrete
changes such as user equipment (UE) movement, tracking data dependencies is a
more efficient computational model than managing a continuous timeline of
events. This design is realised through a \textit{smart update} mechanism, a
practical implementation of lazy evaluation where expressions are not evaluated
until their results are explicitly needed. This approach delivers a measured
speed-up factor of at least 2, compared to a full system recalculation in
typical mobility scenarios.

The basic mathematical blocks, each represented in the code by a python class,
with a NumPy array holding the data, are as follows. All geometric calculations in CRRM are in three dimensions, partly because some of the pathloss models require this, but also to allow modelling scenarios such as tall buildings. Further details are in the docstrings for each class or function.

\begin{itemize}\itemsep0em
\item $U$: a 3-column array in which row $i$ is the position of UE$_i$. 
\item $C$: a 3-column array in which row $j$ is the position of Cell$_j$. 
\item $D$: Distance matrix, defined by $D_{ij}=|\!|u_i-c_j|\!|$. This class computes both 2d and 3d distances, and also angles for use in antenna radiations patterns.
\item $P$: $p_{jk}$ represent the current transmit power of Cell$_j$ in subband $k$.
\item $a$: attachment vector, with the meaning that UE$_i$ is attached to Cell$_{a_j}$. 
\item $G$: gain matrix, defined as some function $g$ of the distance matrix as $G=g(D)$, in which $g$ is defined by a pathloss model. It satisfies $\implies0\leqslant G<1$.
\item $R$: RSRP (received signal reference power) matrix, in which $R_{ijk}=p_{jk} G_{ij}$, with $k$ indexing the subband.
\item $w$: the wanted signal vector, computed from $w_{ik}=R_{i,a_i,k}$.
\item $u$: the unwanted interference vector defined by $u_{ik}=\sum_j R_{ijk}-w_{ik}$. 
\item $\gamma$: the signal plus interfernce to noise ratio (SINR) vector $\gamma=\frac{w}{\sigma^2+u}$, in which division is element-wise. 
\item CQI (channel quality index): this is computed from the SINR in dB, using a look-up table. The values are in the range $[0,15]$.
\item Shannon:  this block computes channel capacity, including for MIMO, using information theory. It is an upper bound on channel throughput.
\item MCS (modulation and coding scheme index): this is a scaled version of CQI. The values are in the range $[0,28]$. These values are mapped onto actual data rates for the different modulation and coding schemes using standard 3GPP tables.
\item Resource allocation. See Section~\ref{Resource Allocation}.
\item Throughput: the final output is computed by combining MCS, resource allocation, and bandwidth.
\end{itemize}

\begin{figure}[p]
\centering
\includegraphics[width=\textwidth]{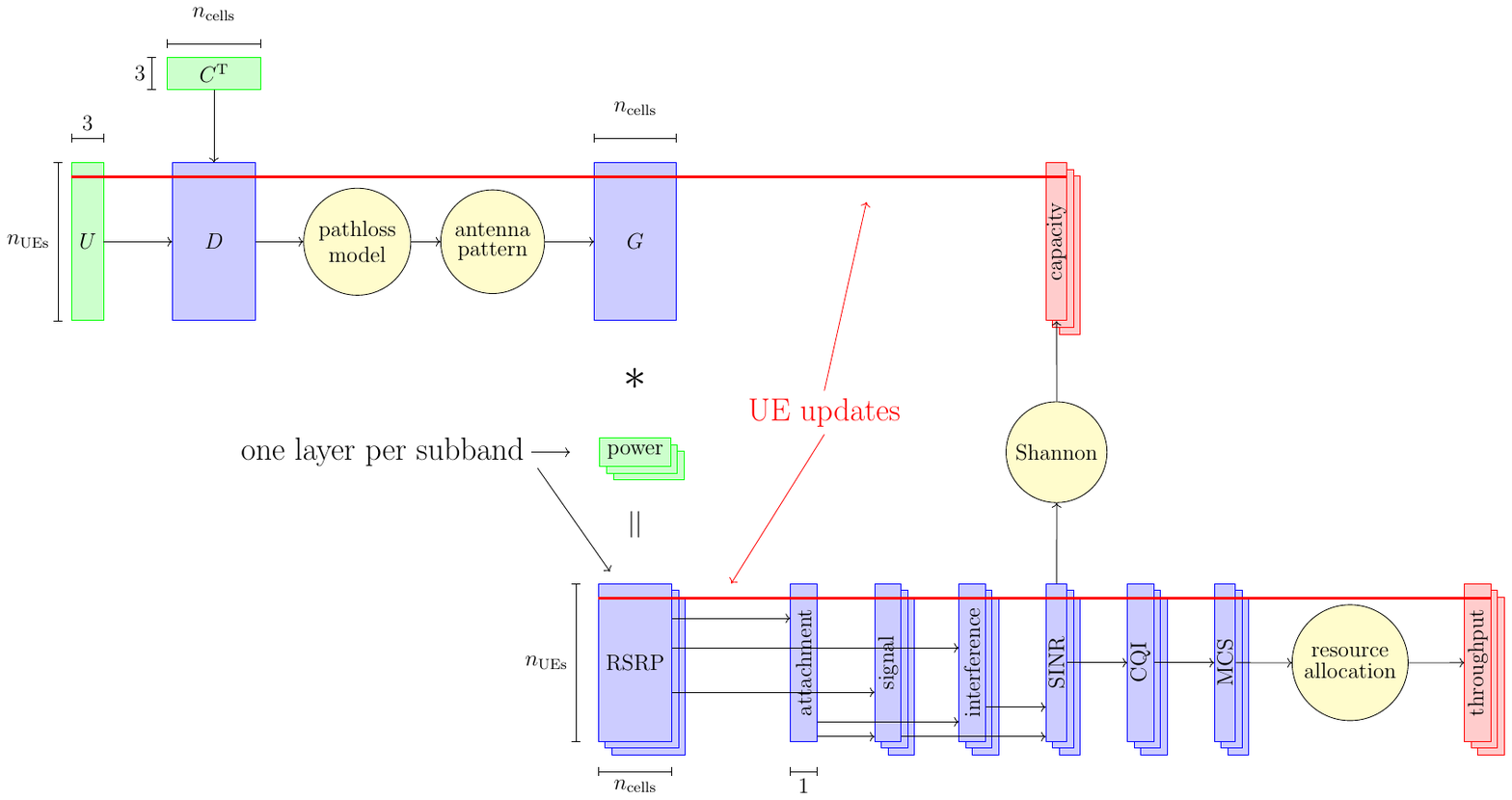}
\caption{The CRRM computational data flow, with the shapes of the arrays
indicated. Data propagates from root nodes like $U$ (representing UE
locations), $C$, and $P$ (power), through a series of dependent blocks, each
representing a distinct mathematical operation. The stacked blocks for RSRP,
SINR, etc.\ illustrate the handling of multiple subbands. The horizontal red stripe indicates the rows of all the arrays requiring to be updated if a UE moves. Python advanced indexing is used to perform multiple UE move updates in one operation.}
\label{fig:block_diagram}
\end{figure} 

The flow of data and the mathematical dependencies between blocks are shown in
Figure~\ref{fig:block_diagram}. The simulation pipeline is a sequence of
transformations beginning with the root inputs, until the final outputs of
capacity of throughput are reached.  This compute-on-demand mechanism is
orchestrated by components within the internal \texttt{\_Node} base class: a
boolean \texttt{up\_to\_date} flag and lists for dependencies
(\textit{watchees}) and dependents (\textit{watchers}). 

The process consists of two distinct phases. First, an invalidation phase
occurs when a root node's data is altered, for example when a UE moves. This
triggers the \texttt{flood\_out\_of\_date()} method, which recursively
traverses the watchers' lists to set the \texttt{up\_to\_date} flag of all
downstream nodes to False. This cascade efficiently validates the specific path
of computations affected by the change without performing any new calculations.
Second, a recursive update phase is initiated when a final result is requested,
for example by \texttt{get\_UE\_throughputs()}. This calls the
\texttt{update()} method on the terminal node, which first checks its
\texttt{up\_to\_date} flag. If False, it recursively calls \texttt{update()} on
each of its watchees.

This process continues up the dependency chain until it reaches nodes that are
already up-to-date. As the recursion unwinds, each node, now guaranteed that
its inputs are current, executes its specific \texttt{update\_data()} method to
re-compute its state.

A key strength of this architecture is its high degree of modularity, realised
by a pluggable physics engine that uses the strategy design pattern. At
initialisation, the \texttt{CRRM\_parameters} class accepts a pathloss model
name as a string (e.g., \texttt{RMa}). The main simulator class then
instantiates the corresponding Python class (e.g., \texttt{RMa\_pathloss}) and
assigns its \texttt{get\_pathgain} method to a generic
\texttt{pathgain\_function} callable. This function is subsequently used by the
\texttt{Gain\_matrix} node during its computation. This design cleanly
separates the simulation's core dependency logic from the specific mathematical
models of radio propagation. The result is a highly extensible system where new
models can be added by creating a new class with the required interface, a
robustness demonstrated by the existing suite of 3GPP-compliant models (RMa,
UMa, UMi, InH).

The decision to implement CRRM in pure Python with NumPy, rather than a
compiled language like \cpp, was a strategic one. The explicit design goal was
to create a tool optimized for usability and direct integration with the AI and
machine learning research ecosystem, which is overwhelmingly dominated by
Python. This choice targets researchers who require a realistic simulation
environment for tasks like reinforcement learning but who cannot afford the
steep learning curve and integration friction of \cpp-based tools. The
compute-on-demand architecture's performance is most effective in typical
low-to-medium mobility scenarios; a stress test where 100\% of UEs move each
time step would logically diminish the gains from lazy evaluation, defining the
operational boundaries of the design. This positions CRRM as a tool that
strategically balances fidelity and raw speed against accessibility.

\section{Features and capabilities}

The utility of the CRRM simulator is defined by its implemented features, which enable the analysis of specific 5G system behaviours. The following sections detail the propagation models, antenna configurations, and resource management algorithms that have been implemented and verified.

\subsection{Propagation environments}

\begin{figure}[ht]
\centering
\includegraphics[width=0.8\textwidth]{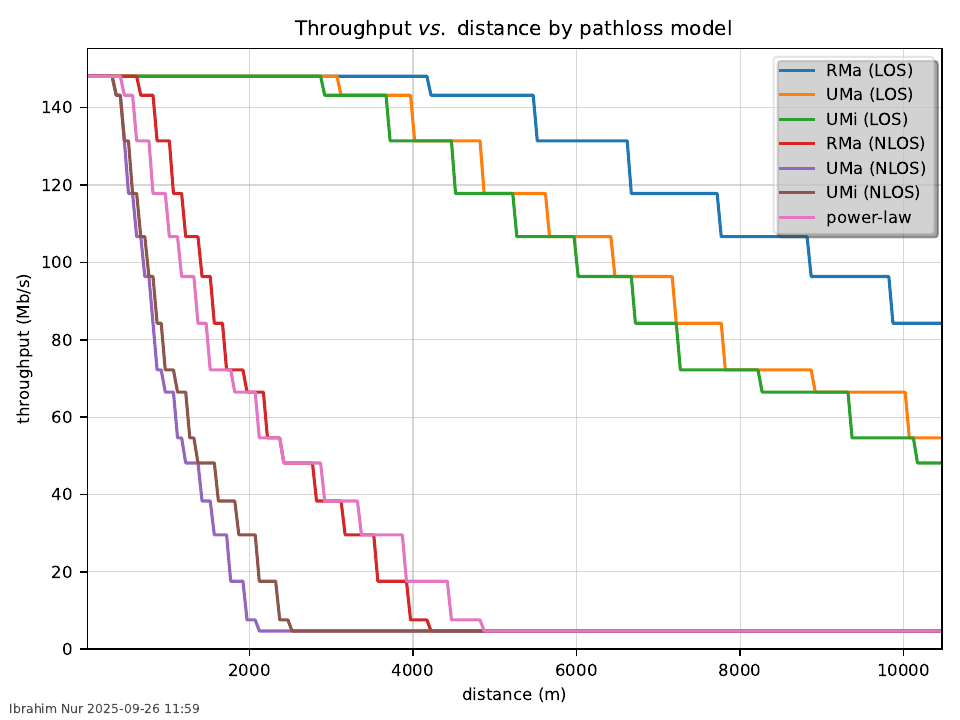}
\caption{Throughput as a function of distance for the RMa, UMa, UMi, and power-law pathloss models. The simulation captures the distinct decay characteristics of each propagation environment.}
\label{fig:pathloss_comparison}
\end{figure}

CRRM includes implementations of the standard 3GPP TR 38.901 pathloss models \cite{3GPP_TR38901} for Rural Macrocell (RMa), Urban Macrocell (UMa), Urban Microcell (UMi), and Indoor Hotspot (InH). This allows the simulator to model the signal propagation characteristics specific to each of these scenarios. Figure~\ref{fig:pathloss_comparison} demonstrates the direct impact of model selection on UE performance, plotting the calculated throughput as a single UE moves radially away from a base station. The simulation shows that for a UE at a distance of 2000 metres in NLOS conditions, the RMa model predicts a throughput of approximately 67 Mb/s, whereas the more obstructive UMa model predicts a throughput of less than 10 Mb/s under the same conditions.

The implementation of the RMa model serves as a case study in engineering
trade-offs. Three variants are provided: \texttt{RMa\_pathloss}, which performs
a full dynamic calculation for any given UE and BS heights;
\texttt{RMa\_pathloss\_constant\_height}, a faster version for scenarios where
all antennas have fixed heights; and 
\texttt{RMa\_pathloss\_discretised}, an optimized model that gains a
significant runtime improvement by using a pre-calculated lookup table of
coefficients for discrete antenna heights. This discretised model achieves its
speed with a root-mean-square error (RMSE) of just 0.16 dB in non-line-of-sight
scenarios when compared to the full model.

\subsection{Advanced antenna and cell modelling}

\begin{figure}[ht]
\centering
\includegraphics[width=0.8\textwidth]{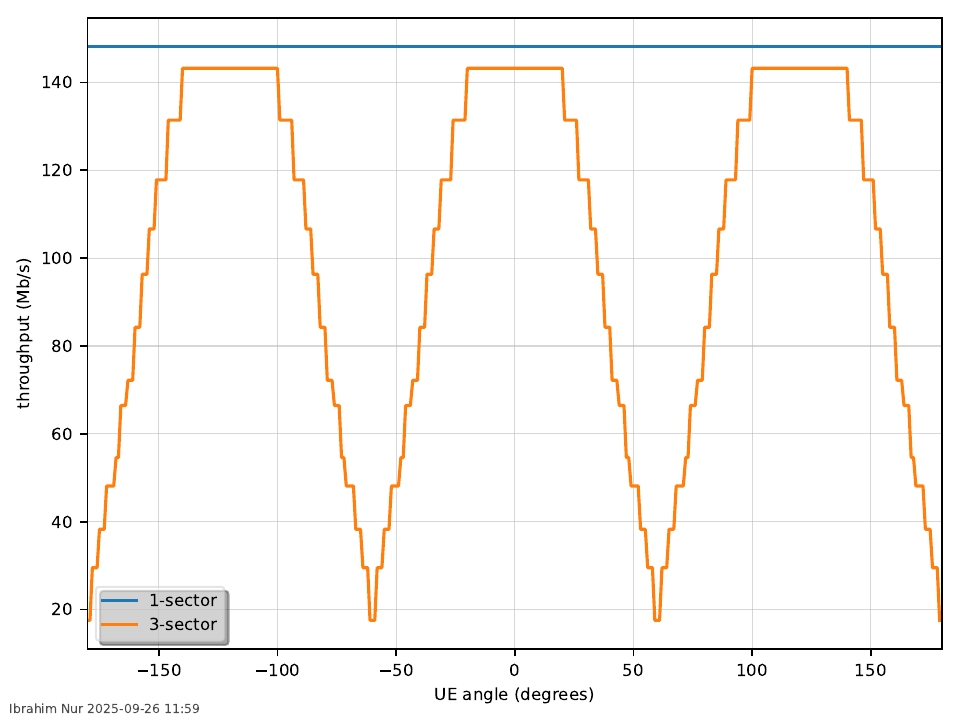}
\caption{Comparison of UE throughput for 1-sector (omnidirectional) and 3-sector antenna configurations. The plot shows the impact of the 3-sector antenna gain pattern as a UE moves in a circle around the base station.}
\label{fig:antenna_pattern}
\end{figure}

To model realistic cell deployments, CRRM implements the 3GPP horizontal
antenna pattern in the \texttt{Antenna\_gain} class. This model allows for the
simulation of sectored base stations, where antenna gain is a function of the
angle relative to the sector's boresight. The model uses the standard
parameters of a 65-degree half-power beamwidth and a maximum attenuation of
30dB. The effect of this sectorisation is demonstrated in
Figure~\ref{fig:antenna_pattern}, which plots the throughput of a UE moving in
a 360-degree circle around a single base station. For a 1-sector
(omnidirectional) configuration, the throughput remains constant. For a
3-sector configuration, the plot clearly shows three distinct lobes where
throughput peaks when the UE is aligned with a sector's centre and drops
significantly in the crossover regions. This confirms the model's ability to
capture the performance impact of directional antenna gains.

\subsection{Radio resource management}

The simulator provides specific mechanisms to study radio resource management strategies. The references here to CRRM example codes are those in the root folder of the CRRM distribution code.

\subsubsection{Subbands}

CRRM supports the division of the channel bandwidth into multiple subbands.
This feature is essential for modelling frequency-domain interference
management techniques. The \texttt{power\_matrix} parameter allows the
transmission power of each cell to be defined on a per-subband basis. A
worst-case interference scenario is demonstrated in the CRRM example code 06,
in which a single UE is placed equidistantly between two cells. When both cells
transmit on the same single subband, the UE experiences an SINR of 0dB. By
reconfiguring the simulation to use two subbands and setting the
\texttt{power\_matrix} so that each cell transmits on a separate subband, the
interference is eliminated and the UE's SINR on its serving subband improves to
20dB. 

\subsubsection{Resource Allocation}
\label{Resource Allocation}

\begin{figure}[ht]
\centering
\includegraphics[width=0.8\textwidth]{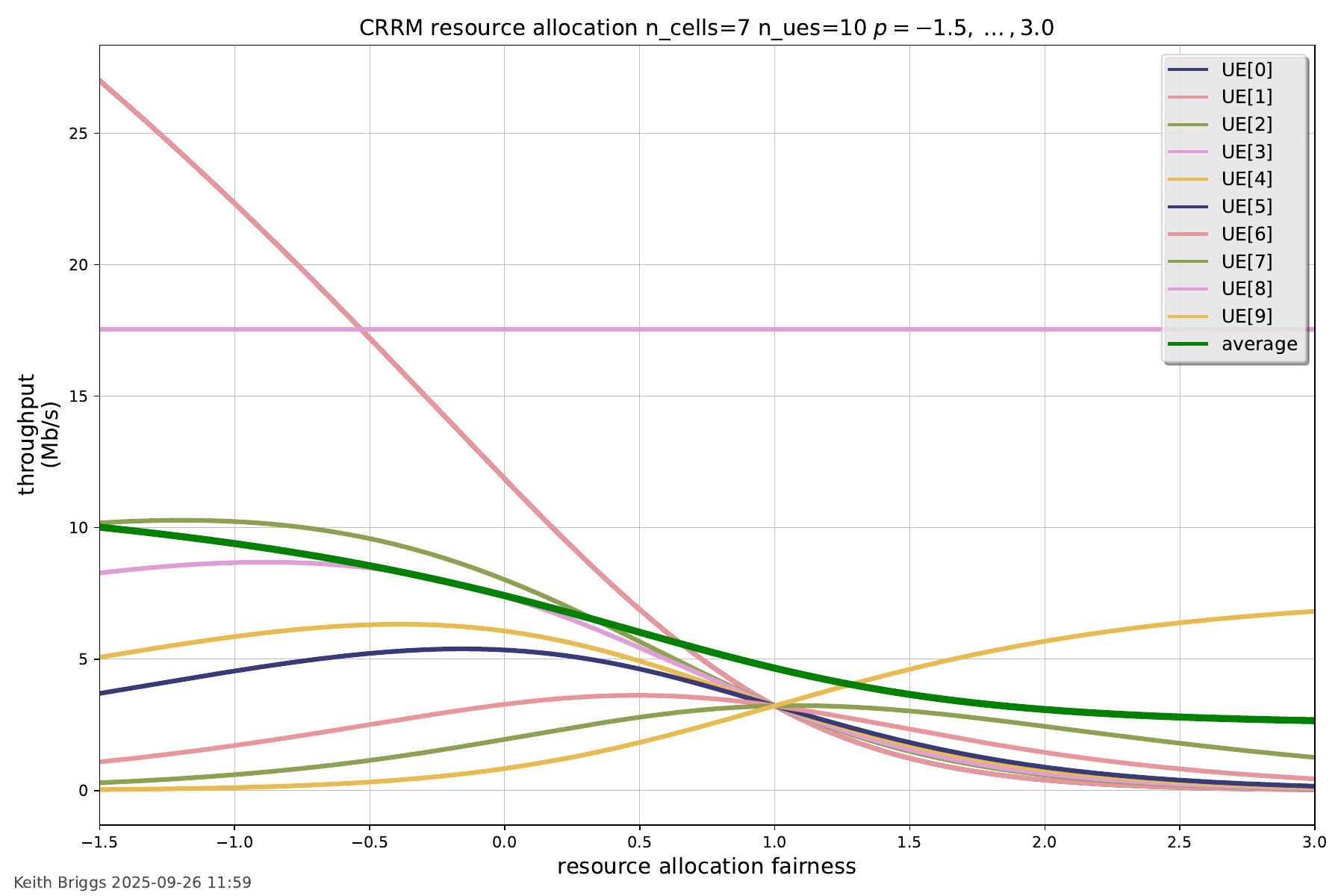}
\caption{UE throughput as a function of the resource allocation fairness parameter, $p$. The plot illustrates how throughput is redistributed from high-spectral-efficiency users to low-spectral-efficiency users as $p$ increases.}
\label{fig:resource_allocation}
\end{figure}

The simulator implements a tunable heuristic for allocating resources among UEs
attached to the same cell. The model is contained within the
\texttt{Throughput} class and calculates the throughput for user $i$,
$T_i$, based on its spectral efficiency, $S_i$, using the formula $T_i = a
S_i^{1-p}$. Here, $p$ is the tunable fairness parameter and $a$ is a
cell-specific normalisation constant. The parameter $p$ directly controls the
allocation strategy: $p=0$ corresponds to proportional fair scheduling where
$T_i \propto S_i$, while $p=1$ results in equal throughput for all users on
the cell. Figure~\ref{fig:resource_allocation} shows the result of a parameter
sweep from CRRM example code 03. It plots the
throughput of multiple UEs as $p$ is varied, demonstrating how the simulator
can be used to analyse the system-level trade-offs between maximising cell
capacity by favouring strong users and ensuring service fairness for weak,
cell-edge users.

\section{Validation and performance}

A simulator's utility is contingent on two factors: the correctness of its results and its computational performance. This section presents the tests conducted to verify CRRM's physical layer calculations against established theory and to quantify the efficiency of its core compute-on-demand architecture.

\subsection{Validation of the core engine}

To validate the correctness of the simulator's interference calculations, a
large-scale network was simulated and its output compared against a known
analytical result from stochastic geometry \parencite{Haenggi2013} The
experiment, contained in CRRM example 12 defines a network of 10,000 base
stations and 1,000 UEs, with their locations distributed according to a Poisson
Point Process (PPP). The simulation was configured to use a simple power-law
pathloss model with an exponent of $3.5$, and thermal noise was set to zero
($\sigma^2 = 0$) to remove the noise component from the the signal to
interference plus noise ratio.  The simulation was run with Rayleigh
fading, as the analytical result assumes.

\begin{figure}[ht]
\centering
\includegraphics[width=0.7\textwidth]{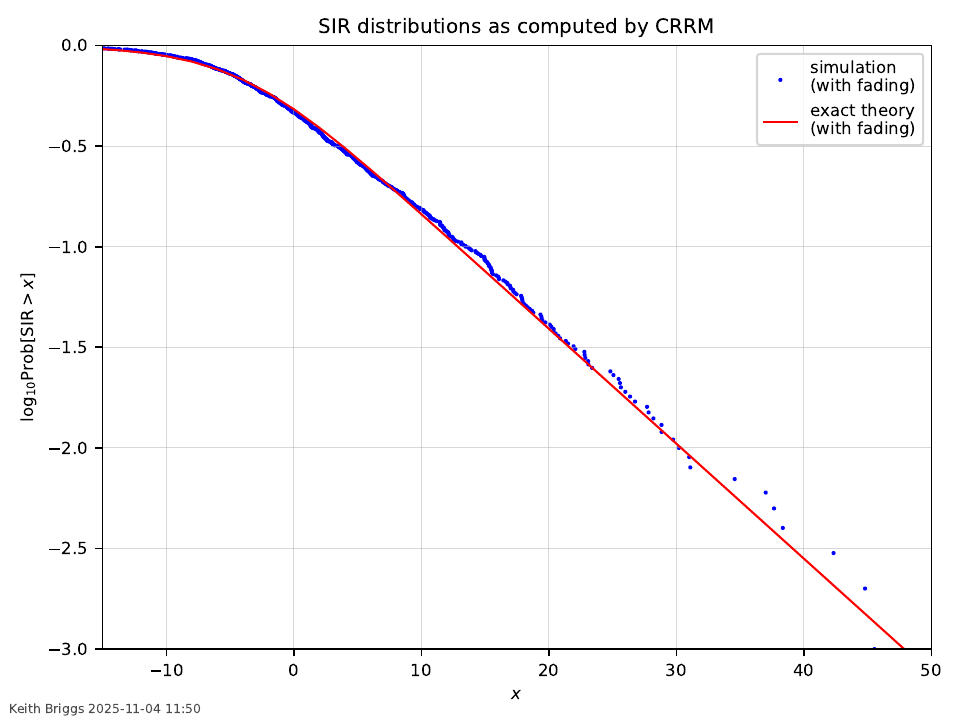}
\caption{Validation of the simulated SIR distribution. The complementary cumulative distribution from a 1000-UE PPP simulation (dots) shows excellent agreement with the exact analytical theory (solid line).}
\label{fig:sinr_cdf}
\end{figure}

The results are shown in Figure~\ref{fig:sinr_cdf}. The plot shows the
complementary cumulative distribution function of the SIR values
produced by the simulation. This empirical distribution is plotted alongside
the exact, theoretical distribution for a PPP network with the same pathloss exponent.
The close agreement between the simulated data points and the solid theoretical
curve provides strong evidence for the correct implementation of the
simulator's core computational engine, from distance calculations through to
the final aggregation of interference.

\subsection{Performance of the compute-on-demand architecture}

The central performance claim of CRRM is that its compute-on-demand
architecture is more efficient than a full system recalculation in typical
mobility scenarios. This claim can be tested with CRRM example 13. The
experiment simulates a network for a fixed number of time steps, where in each
step a fraction of the UEs (10\%) are moved randomly. The test is executed
twice: once with the smart update mechanism enabled, and once with it
disabled, forcing a recalculation of all metrics for all UEs in every step.

The test script first confirms that the final SINR and spectral efficiency
results from both the smart and non-smart runs are numerically identical,
verifying the correctness of the lazy evaluation logic. The primary result is
the measurement of wall-clock time for each run. The test demonstrates that the
smart update mechanism yields a speed-up factor of about 2. This performance
gain is a direct result of the architecture avoiding redundant computations for
the 90\% of UEs that remain static in each time step, thereby confirming the
effectiveness of the design.

\section{Conclusion}

This work has presented the Cellular Radio Reference Model (CRRM), a 5G
system-level simulator designed specifically to address the usability gap for
researchers in the machine learning domain. Its core contribution is a
compute-on-demand architecture, a departure from traditional discrete-event
schedulers, which confines computation to only those network parameters
affected by a state change. This design was shown to yield a speed-up factor of
about 2 in typical 10\% mobility scenarios while producing numerically
identical results to a full system recalculation. The simulator's scientific
credibility was established by validating its SIR calculations for a Poisson
Point Process network against exact analytical theory, where the results showed
excellent agreement. This validated engine is complemented by a comprehensive
feature set, including 3GPP-compliant propagation models for RMa, UMa, UMi, and
InH environments, multi-sector antenna patterns, and mechanisms for analysing
radio resource management.



CRRM successfully fulfils its primary design goal: to provide an accessible,
high-performance simulation environment within the Python ecosystem. It is not
intended to replace complex, full-stack \cpp\ simulators but to serve as a
complementary tool for a specific research community. By balancing physical
layer fidelity with computational efficiency and ease of integration, CRRM
provides a tool that is sufficiently realistic for system-level experiments on
mobility and resource management, while prioritising the rapid prototyping and
direct framework compatibility required for modern, AI-driven research. It
therefore represents a valuable asset for accelerating future work at the
intersection of wireless communications and machine learning.

The potential business impact of this type of simulation is the provision of a
safe, offline environment to test and validate new automation algorithms before
they are considered for deployment on the live, multi-billion pound network.
CRRM’s specific features, such as the per-subband \texttt{power\_matrix} and
the tunable resource allocation fairness parameter, $p$, give engineers the
tools to quantify the direct trade-offs between network-wide resource use and
end-user throughput. This allows for the modelling of policies aimed at
increasing spectral efficiency or reducing total power consumption. The
simulator can therefore produce the quantitative data required to support
network configuration changes that may lead to multi-million pound reductions
in operational expenditure. 

\section{Access to code}

\begin{itemize}\itemsep0em

\item The project homepage is \url{https://keithbriggs.info/crrm.html}.

\item The package can be installed with \texttt{pip}, as described on
\href{https://pypi.org/project/CRRM/}{pypi}. 

\item The full documentation is on
\href{https://crrm-20.readthedocs.io/en/latest/}{readthedocs}.

\item The open-source code is hosted on
\href{https://github.com/keithbriggs/CRRM-2.0}{github}.

\end{itemize}

\printbibliography

\end{document}